\documentclass[prb,floatfix,showpacs,superscriptaddress,longbibliography,twocolumn]{revtex4-1}
\usepackage[USenglish]{babel}
\usepackage{amsmath,amssymb,amsfonts}
\usepackage{epsfig}
\usepackage{graphicx}
\usepackage{subfigure}
\usepackage{import}

\usepackage[colorlinks=true,linkcolor=blue,citecolor=red]{hyperref}

\newcommand{\eq}[2]
{
  \begin{equation}
    #1
    \label{#2}
  \end{equation}
}

\newcommand{\equ}[1]
{Eq.~(\ref{#1})}

\newcommand{\figu}[1]
{Fig.\ref{#1}}

\newcommand{\tddpam}{$t_{dd}$-PAM }
\newcommand{\pam}{$t_{dd}$-PAM }

\def\bcen{\begin{center}}
\def\ecen{\end{center}}

\def\a{\alpha}             \def\d{\delta} 
\def\e{\varepsilon}          
              
                    \def\s{\sigma}

\def\GG{{\cal G}} 
\def\ZZ{{\cal Z}}

\def\eg{\mbox{\it e.g.\ }}  \def\ie{\mbox{\it i.e.\ }}
\def\=={\equiv}

\def\qed{\raise1pt\hbox{\vrule height5pt width5pt depth0pt}}
\def\iome{i\omega_n}

\def\cG0{{\cal G}_0} 
\def\cG{{\cal G}}

\def\bra{\langle} \def\ket{\rangle}
\def\ka{{\bf k}}

  \def\Im{\mbox{Im}}
\def\ie{\hbox{\it i.e.\ }} \def\eg{\mbox{\it e.g.\ }}
\def\ie{\mbox{\it i.e.\ }} \def\=={\equiv}
 
\def\Im{{\rm Im}}  
 \def\ep0{\epsilon_{p}} \def\ed0{\epsilon_{d}}
\def\tpd{t_{pd}}

\begin{document}
\title{Local moment dynamics and screening effects \\
in doped charge-transfer insulators}

\author{A.~Amaricci} 
\affiliation{Democritos National Simulation Center, 
Consiglio Nazionale delle Ricerche, 
Istituto Officina dei Materiali (IOM) and 
Scuola Internazionale Superiore di Studi Avanzati (SISSA), 
Via Bonomea 265, 34136 Trieste, Italy}
\author{N.~Parragh}
\affiliation{Institute for Theoretical Physics and Astrophysics, 
University of W\"urzburg, D-97074 W\"urzburg, Germany}
\author{M.~Capone}
\affiliation{Democritos National Simulation Center, 
Consiglio Nazionale delle Ricerche, 
Istituto Officina dei Materiali (IOM) and 
Scuola Internazionale Superiore di Studi Avanzati (SISSA), 
Via Bonomea 265, 34136 Trieste, Italy}
\author{G.~Sangiovanni}
\affiliation{Institute for Theoretical Physics and Astrophysics, 
University of W\"urzburg, D-97074 W\"urzburg, Germany}

\date{\today}

\begin{abstract}
By means of Dynamical Mean-Field Theory we investigate the spin
response function of a model for correlated materials with $d$- or
$f$-electrons hybridized  with more delocalized ligand orbitals.
We point out the existence of two different processes responsible for the
dynamical screening of local moments of the correlated electrons. 
Studying the local spin susceptibility we identify the contribution of the 
``direct'' magnetic exchange and of an ``indirect'' one mediated by
the itinerant uncorrelated orbitals.
In addition, we characterize the nature of the dynamical screening
processes in terms of different classes of diagrams in the
hybridization-expansion contributing to the density-matrix. 
Our analysis suggests possible ways of estimating the relative
importance of these two classes of screening processes in realistic
calculations for correlated materials.
\end{abstract}


\maketitle

\section{Introduction}
The physics of strongly correlated systems can be identified with the
process that turns delocalized electrons into localized magnetic
moments. 
In $d$- and $f$-electron materials, such as transition-metal oxides
(TMO) and heavy fermions (HF), the confined nature of the correlated
orbitals  gives rise to large spin moments localized over short time
scales at each lattice site.
In the atomic limit, in which the system is described as a collection of 
disconnected atoms, such local spins assume the largest value 
allowed by the atomic configuration, and show no dynamics.
On the other hand, in the presence of a fraction of itinerant electrons 
(\ie away from the extreme case of the atomic limit), the instantaneous
value of the local spins gets reduced and a dynamics emerges as an effect of 
the screening processes.

In TMO sizeable local magnetic moments from $d$-orbital electrons 
have been revealed by inelastic neutron scattering (INS) and X-ray absorption
spectroscopy (RIXS)\cite{Ament2011} in materials like,\eg  high-T$_c$
cuprates\cite{Braicovich2010,Tacon2011},
cobaltate\cite{Kroll2006PRB,Lang2008PRB} and iron-based
superconductors \cite{Chi2009PRL,Liu2012NP,Zhou2013NC}.
The fingerprints of the individual screening processes are hard to extract 
from the instantaneous value of the local moments. 
For instance, only indirect information about the nature of the
screening of the local moments can be gained from the
temperature dependence of the local spin susceptibility\cite{Raas2009,Grete2011}. 
The spin dynamics can instead be a much more sensitive tool to
diagnose the  physical effects of such screening processes \cite{Hansmann2010PRL,Toschi2012PRB}.  

From a general point of view one expects two 
processes to be responsible for the screening of the local moments: 
$i$) processes involving a {\it direct} hopping between correlated
electrons, and 
$ii$) processes involving the {\it hybridization} with more itinerant,
\eg ligand $p$-, orbitals.
In the oxides, one associates processes of type $i$) with the
super-exchange mechanism which is captured by a Hubbard-like
description including only correlated $d$-orbitals.  
This ``$d$-only'' description is good whenever the $d$-manifold is
very well isolated  from the ligand $p$-bands.
Processes of type $ii$) can instead be viewed as an additional
screening channel, active if the system can gain delocalization
energy upon hybridizing with the bath of itinerant electrons. 

In the present paper we study the local moment dynamics and the screening effects
addressing questions such as:
``how can we distinguish between type-$i$) and type-$ii$) 
contributions  to local spin susceptibility?'', or 
``which features would we expect to see in a INS or RIXS
experiment if the screening is dominated by
hybridization processes?''
In order to do this we focus our attention on doped
``charge-transfer'' insulators \cite{Imada1998RMP} (CTI), \eg Cuprates, as a paradigmatic
example of Mott systems in which the effects of the hybridization are crucial.

We therefore consider a generic, yet simple, model for a CTI in which
the doping and the relative importance between the two screening channels 
can be easily tuned. 
By using Dynamical Mean-Field Theory (DMFT)\cite{Georges1996RMP} we study the 
local dynamical spin response function. 
We point out the existence of two distinct features in the local spin
susceptibility associated to processes of the two different types and we
numerically characterize their nature in a clear way in terms of
different classes of diagrams contributing to the
density-matrix. 

The structure of the paper is as follow: in Sec.~\ref{secModel} we
introduce the theoretical model and briefly discuss its numerical
solutions within DMFT. In Sec.~\ref{secChi} we discuss the results for
the local moment dynamics. In Sec.~\ref{secDiagrams} we characterize the different
features in the spin susceptibility in terms of distinct class of
diagrams in the strong-coupling expansion. Finally, section~\ref{secConclusions}
contains concluding remarks.

\section{Model}\label{secModel}
We  consider a generalized
periodic Anderson  model (\pam) \cite{Medici2005PRL} describing a 
wide-band of conduction electrons,  hybridizing with a 
narrow-band of strongly interacting electrons: 
\begin{equation}
\begin{split}
H=&\sum_{\ka\s}\e_p(\ka)p^+_{\ka\s}p_{\ka\s} + \sum_{\ka\s}\e_d(\ka)d^+_{\ka\s}d_{\ka\s}+\\ 
&\tpd\sum_{i\s}\left(d^+_{i\s}p_{i\s} + p^+_{i\s}d_{i\s}\right) 
+ U\sum_i d^+_{i\uparrow}d^+_{i\downarrow}d_{i\uparrow}d_{i\downarrow}\\
\end{split}
\label{Hpam}
\end{equation}

The operators $p_{i\s}$ ($p^+_{i\s}$) destruct (create) electrons 
in the conduction band with spin-$\s$ with dispersion
$\e_p(\ka)=\ep0-2t_{pp}[\cos(k_x)+\cos(k_y)]$. 
Similarly, $d_{i\s}$ ($d^+_{i\s}$)  destruct (create) electrons in the
narrow-band with spin-$\s$ and dispersion $\e_d(\ka) = \ed0 -
2\a t_{pp}[\cos(k_x)+\cos(k_y)]$, where $\a\in[0,1)$ denotes the
bandwidth ratio.
The two orbital electrons hybridize with a local amplitude $\tpd$. 
The last term in Hamiltonian (\ref{Hpam}) indicates the strong local Coulomb
interaction $U$ experienced by the $d$-electrons. 

In the following, we fix the energy unit to the half-bandwidth of
the conduction electrons $D=4t_{pp}=1$. In addition we shall set the bandwidth ratio
to $\alpha=0.25$ and $\ed0=0$.
The energy separation between the centers of the two bands $\Delta=\ep0-\ed0$ denotes the
charge-transfer energy. Finally, we will drop any reference to the
spin index, as we focus on the paramagnetic state, where the local moments 
are not ordered.

The model Hamiltonian (\ref{Hpam}) interpolates between the Hubbard 
model (HM) for the correlated $d$-electrons ($\tpd=0$, $\a\neq0$), 
and the more usual periodic Anderson model ($\a=0$,
$\tpd\neq0$),  describing non-dispersive correlated electrons
hybridized with a wide-band\cite{Jarrell1993PRL,Jarrell1995PRB,Pruschke2000,Sordi2007PRL}. 

We solve the \pam using Dynamical Mean Field Theory
(DMFT)\cite{Georges1996RMP}.  The DMFT allows us to study the local
screening and the spin dynamics in a fully non-perturbative way. 
Within DMFT, the lattice model (\ref{Hpam}) is mapped onto an
effective impurity problem for a single $d$-orbital, supplemented by a
self-consistency condition for the local {\it Weiss Field} (WF). 
The WF $\GG_{0dd}^{-1}$ is calculated by isolating
the $dd$-element of the interacting local Green's function, as in
general DMFT schemes with enlarged basis-sets \cite{Lechermann2006PRB,Han2011PRL,Parragh2013a,Haule2013a}. 
We solve the associated effective impurity problem using 
exact-diagonalization (ED)\cite{Caffarel1994a,Capone2007PRB,Weber2012PRB}  and
hybridization-expansion continuous-time quantum
Monte Carlo (CTQMC) methods\cite{Werner2006PRL,Haule2007PRB,Lauchli2009,Parragh2012PRB}. 

Within ED the local WF must be represented in terms of a discretized hybridization function:
$\Delta_{dd}(\iome) = \sum_{l=1}^{N_b}{V_l^2}/{(\iome-\e_l)}$  
using $N_b$ auxiliary energy levels. The parameters $\e_l$ and $V_l$ describe,
respectively, the local energy and the hybridization between the impurity 
and the $l^{\rm th}$ bath level.
All the ED calculations are performed using $N_b=6$. 

The  spin susceptibility $\chi_\mathrm{spin}$ is defined as the imaginary part
of the dynamical response function:
\eq{
\chi_\mathrm{spin}(\omega)=i \int dt e^{i\omega t} 
\theta(t) \bra [\hat{S}^{(d)}_z(t),\hat{S}^{(d)}_z(0)]\ket 
%\mathrm{Tr}\bra [\hat{S}^{(d)}_z(t),\hat{S}^{(d)}_z(0)]\ket \theta(t)
}{chidef}
where $\hat{S}^{(d)}_z$ is the $z$-component of the spin operator on
the $d$-site  and $[ \, , \, ]$ denotes the commutator.
In ED the spin susceptibility is evaluated using the spectral decomposition:
\eq{
\begin{split}
 \chi''_\mathrm{spin}(\omega) = &
 \frac{\pi}{\ZZ} \sum_{i,j} | \bra i| \hat{S}^{(d)}_z |j\ket |^2  
 (e^{-\beta E_j} + \\
& -e^{-\beta E_i}) \delta \left( \omega -(E_i-E_j) \right),
\end{split}
}{Imchi}
where $\ZZ$ is the partition function.
\noindent

The hybridization-expansion CTQMC method provides a (statistically)
exact solution of  the DMFT equations.  
Indeed, we tested the agreement between ED and CTQMC calculations
finding very satisfactory results for both local and dynamical quantities. 
As we will show, CTQMC permits to investigate the diagrams
contributing to the local screening processes. 
Since this is done using a (infinite series) perturbation-expansion
language, it turns out to give useful information about the physics
involved in the screening of the local moment.

%%%%%%%%%%%%%%%%%%%%%%%%%%%%%%%%%%%%%%%%%%%%%%%%%%
% FIGURE 1
\begin{figure}[t]
 \includegraphics[width=0.95\linewidth]{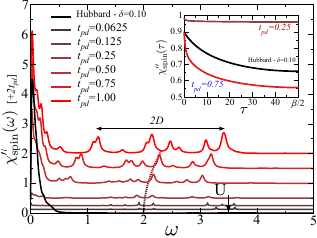}
 \caption{(Color online) Spin susceptibility
   $\chi''_\mathrm{spin}(\omega)$ on the real-axis for increasing values of $\tpd$.  
   The different curves are shifted along the $y$-axis by $2\tpd$ for
   better comparison. 
   Data are for doping value $\delta=0.10$. 
   Dotted line is the shift in the position of the $p$-band from
   increasing hybridized band repulsion.
   Inset: imaginary-time spin susceptibility $\chi''_\mathrm{spin}(\tau)$
   for a fixed doping $\delta=0.10$ for $\tpd=0.25$ and $0.75$.
   The black curve corresponds to the Hubbard model result with doping $\d=0.10$.
}
\label{fig1} 
\end{figure}
%%%%%%%%%%%%%%%%%%%%%%%%%%%%%%%%%%%%%%%%%%%%%%%%%%

\section{Spin susceptibility}\label{secChi}
As discussed in the Introduction, we focus on doped CTI. 
By definition this means that the hole doping involves mainly the
$p$-band. 
This marks a strong difference with a description of pure
$d$-electrons, where the insulator has a Mott-Hubbard character.
In order to study the differences in the spin susceptibility and in
other observables induced by the hybridization $\tpd$, we want to be
able to compare solutions with both finite $\tpd$ and finite doping to
a \emph{doped} Hubbard model. 
However, we can not recover the latter in the limit of vanishing
$\tpd$ of our model, as this tends towards a \emph{half-filled},
``$d$-only''  Mott insulator. 
Hence, we shall complement our calculations by solving the Hubbard
model  for a given (in principle arbitrary) value of the hole-doping $\delta$. 
We will fix $\delta$ 
%for these ``$d$-only'' Hubbard calculations 
according to physically motivated 
criteria, \eg that the size of the instantaneous spin moment is that of 
the solution with finite $\tpd$ we are comparing to, or, that the
occupation  of the $d$-orbital $\bra n_d \ket$ is the same between the two models.

In order to place the system into the ``charge-transfer'' regime we
consider the model defined in \equ{Hpam} with
$U=3.5$ and $\Delta=-0.5$. In addition we set the
temperature to $T=1/100$.

%%%%%%%%%%%%%%%%%%%%%%%%%%%%%%%%%%%%%%%%%%%%%%%%%%
% FIGURE 2
\begin{figure}[t]
 \includegraphics[width=0.95\linewidth]{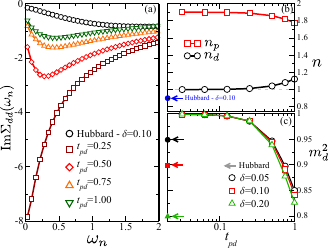}
 \caption{(Color online) (a) Imaginary part of the Matsubara $d$-electron
   self-energy $\Im\Sigma_{dd}$ for increasing hybridization
   amplitude $\tpd$ and doping $\delta=0.10$.
   %The other model parameters are $U=3.5$,
   %$\Delta=-0.5$,  $T=1/100$ and $\delta=0.10$. 
   (b) $p$- and $d$-orbital densities, $\bra n_p\ket$
   and $\bra n_d\ket$ respectively, as a function of $\tpd$. 
   The arrow to filled symbol indicates $\bra n_d\ket$ for the Hubbard
   model with doping $\delta=0.10$.
   (c) $d$-electron local moment $m^2_d=\bra \hat{S}^{(d)}_z{}^2 \ket$ 
   as a function of $\tpd$ and different values of the doping $\delta$. 
   The arrows indicate the Hubbard model results at the same values of
   $\d$.
 }
\label{fig2} 
\end{figure}
%%%%%%%%%%%%%%%%%%%%%%%%%%%%%%%%%%%%%%%%%%%%%%%%%%

The existence of two different screening channels of the local moments 
has a very strong  effect on the spin dynamics. 
This is illustrated in Fig.\ref{fig1}, where results for the local
spin  susceptibility $\chi_\mathrm{spin}$ (both in $\omega$ and in
$\tau$) are shown  for different values of $\tpd$  and total
occupation  $n=\langle n_d \rangle + \langle n_p \rangle = 2.9$,
as well as for the ``$d$-only'' Hubbard model at $\delta=0.10$.
As it can be seen in the inset of Fig.\ref{fig1} and in
Fig.\ref{fig2}c, where we show the  $d$-electrons local moment 
$m^2_d=\bra \hat{S}^{(d)}_z{}^2 \ket$, the latter calculation yields
the  same instantaneous (\ie $\tau=0$) moment of the case of the
\tddpam  with $\tpd=0.75$.

In order to disentangle the contribution of the screening channels, 
we shall discuss the behavior of $\chi_\mathrm{spin}(\omega)$ of 
Fig.\ref{fig1} in more detail. 
In the ``$d$-only'' Hubbard case the instantaneous local moment 
is dynamically screened by coherent metallic excitations at
the Fermi level, as indicated by the vanishing 
imaginary part of the  self-energy (see Fig.\ref{fig2}a). 
In this regime, the screening process is entirely coming from $d$-$d$
direct exchange with a leading coupling  $J_{dd}\simeq\a^2t^2_{pp}/U$,
as the $p$-electrons are completely decoupled.
Correspondingly, the spin susceptibility $\chi_\mathrm{spin}(\omega)$
is dominated  by low-energy contributions, though weaker high-energy
features  at $\omega\simeq U$, associated to electronic excitations
across  the Hubbard bands, can be detected.

%%%%%%%%%%%%%%%%%%%%%%%%%%%%%%%%%%%%%%%%%%%%%%%%%%
%FIGURE 3
\begin{figure}[t]
 \includegraphics[width=0.95\linewidth]{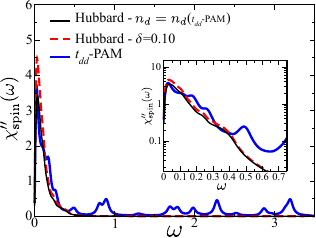}
 \caption{(Color online) Comparison of the spin susceptibility
   $\chi_\mathrm{spin}(\omega)$ of the \tddpam with $\tpd=0.75$ and
   doping $\d=0.10$ (thick line) 
   with two ``$d$-only'' Hubbard model calculations (thin and dashed lines). 
   The thin line shows the case with $d$-electrons occupation 
   $\bra n_d\ket$ equal to the value of $n_d$ in the \tddpam. 
   The dashed line show the case at fixed doping $\d=0.10$.
   Inset: blow-up of the low-frequency behavior from the main panel.
}
\label{fig3}
\end{figure}
%%%%%%%%%%%%%%%%%%%%%%%%%%%%%%%%%%%%%%%%%%%%%%%%%%

If we now consider the extreme case of very large hybridization strength,
\ie $\tpd\!=\!1$, we notice pronounced changes in the spin susceptibility.
First of all the low-frequency part acquires more structures, as
further  underlined in Fig.\ref{fig3} where the case $\tpd\!=\!0.75$ is
directly  compared to two ``$d$-only'' solutions: one for
$\delta\!=\!0.10$ and the other for the same value of $\langle n_d
\rangle$ as in our model. 
Big differences can be seen in the intermediate-to-high frequency region.
The $\chi_\mathrm{spin}(\omega)$ of the $t_{dd}$-PAM acquires there a
significant weight and  several additional peaks are visible. 
As highlighted in Fig.\ref{fig1}, these features extend in a frequency
range with  a width set by the  $p$-electrons bandwidth $2D$, while
their position  scales with $\sqrt{\Delta^2 + 4\tpd^2}$.
The latter corresponds to the correction to the charge-transfer energy
from hybridized bands repulsion, confirming that the
intermediate-to-high lying peaks come from the
hybridization with the more delocalized $p$-orbitals. 

We now focus on the low-frequency region, \ie for $\omega < 1$. A
blow-up is shown in  the inset to Fig.\ref{fig3}.
Evidently, the two ``$d$-only'' calculations both display 
less structures than the $t_{dd}$-PAM (blue line).
This suggests that the low-frequency feature of the spin
susceptibility of the \tddpam has a mixed $d$-$p$ character.
In fact, in this region we expect both the $d$-$d$ screening
processes and the $d$-$p$ ones to be active. To estimate the
order of the screening processes one can consider that in the presence of
finite hybridization the $d$-electrons have an effective hopping of
the order $t_\mathrm{eff}=\a t_{pp} + t_{pd}^2/\Delta$. Then, using a simple
super-exchange argument, we can associate a number of coupling
constants to the different local moments screening processes as follows:
\begin{equation} \label{Js}
J_{dd}\simeq \frac{\alpha^2 t_{pp}^2}{U}\,,\quad 
J_{pd}^{(1)}\simeq \frac{2\a t_{pp}}{\Delta U}\tpd^2\,,\quad
J_{pd}^{(2)}\simeq \frac{\tpd^4}{\Delta^2 U}\,.
\end{equation}

As pointed out before, the first constant describes direct $d$-$d$
processes. The other two describe screening processes involving two or
four hybridizations with non-interacting electrons and, respectively, one or no
direct hopping events.  For large values of $\tpd$ the $J_{pd}^{(1)}$
dominates at  low-energy. On the other hand, for small value of the
hybridizations  $J_{pd}^{(2)}$ becomes smaller than $J_{pd}^{(1)}$ and the associated
exchanges processes dominates at low-frequency.

%%%%%%%%%%%%%%%%%%%%%%%%%%%%%%%%%%%%%%%%%%%%%%%%%%
% FIGURE 4
\begin{figure}[t]
 \includegraphics[width=0.95\linewidth]{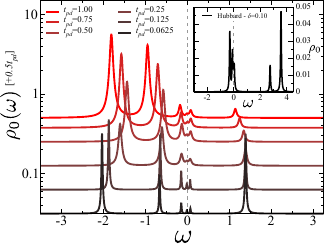}
 \caption{(Color online)  Spectral density of the hybridization
   function  $\rho_0(\omega)=-\Im\Delta_{dd}(\omega)/\pi$ for increasing
   amplitudes $\tpd$.
   Inset: the same quantity for the ``$d$-only'' Hubbard model.
}
\label{fig4} 
\end{figure}
%%%%%%%%%%%%%%%%%%%%%%%%%%%%%%%%%%%%%%%%%%%%%%%%%%

We can attempt to relate the estimates of \equ{Js} to the structures
of  $\chi_\mathrm{spin}(\omega)$ shown in the inset to \figu{fig3}.
For the parameters used, $J_{dd}$ is the smallest coupling
($\mathcal{O}$(10$^{-3}$))  and it can be associated to the
lowest-energy  onset of $\chi_\mathrm{spin}(\omega)$ present in all
three  cases.
$J_{pd}^{(1)}$ assumes the value of 0.04 which roughly corresponds to
the  position of the first deviation (dip) between the \tddpam curve
and  the ``$d$-only'' Hubbard solutions. 
The largest coupling $J_{pd}^{(2)}$ (of the order of 0.4) falls
in the region separating the low-energy structures from the 
intermediate-energy ones, where the largest deviations from the
``$d$-only''  Hubbard start to appear. 

% da qui in poi discutere il limite tpd->0
In the remaining part of this section we discuss the behaviour of the 
spin susceptibility as we decrease $\tpd$ down to very small values.
As shown in Fig.\ref{fig2}b this corresponds to reducing the mixed valence
character of the \tddpam solution. As we mentioned above, the
occupation  $\bra n_d\ket$ of the $d$-orbital approaches 1 from above
while $\bra n_p\ket$ saturates to 1.9 in order to keep the total density to 2.9. 
Concomitantly, the size of the instantaneous moment (see \figu{fig2}c)
increases towards  the atomic value of 1. 
The imaginary part of the self-energy becomes very large at low
frequency, reflecting the strong incoherent character of the solution 
\cite{Tahvildar-Zadeh1997PRB,Burdin2000PRL,Burdin2009PRB,Amaricci2008PRL,Amaricci2012PRB}
(see \figu{fig2}a). 
In this regime of very small $\tpd$ the screening is poor. 
Indeed, the spin susceptibility is very small and essentially featureless.  
A remnant of the low-frequency peak can still be detected for
$\tpd=0.125$ and a structure related to the excitations between the
lower Hubbard band and the (suppressed) spectral density at the Fermi
level, is recognizable at energies of order $U$.

 %%%%%%%%%%%%%%%%%%%%%%%%%%%%%%%%%%%%%%%%%%%%%%%%%%
% FIGURE 5
\begin{figure}[t]
 \includegraphics[width=0.95\linewidth]{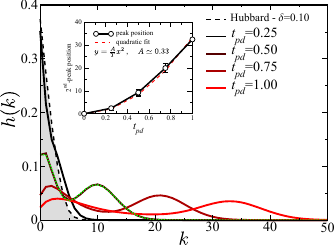}
 \caption{(Color online) Histogram $h(k)$ of the distribution of fermionic
   diagrams  contributing to the local CTQMC trace for different
   expansion order $k$. 
   The histograms are shown for different values of the hybridization $\tpd$ at fixed
   doping $\delta=0.10$. 
   For the Hubbard model case (dashed line) the distribution has one
   single peak.  For the \tddpam ($\tpd\!\neq\!0$) the curves assume a
   bi-modal distribution with both ``low-order'' and ``high-order''
   features. Each of these curves is fitted by a double gaussian (dotted line).  
   Inset: The position of the second peak scales quadratically in 
   $\tpd$. The peak position and error bars are estimated from the mean
   value and standard deviation of the double gaussian fits.
}
\label{fig5} 
\end{figure}
%%%%%%%%%%%%%%%%%%%%%%%%%%%%%%%%%%%%%%%%%%%%%%%%%%

\section{Diagrammatic characterization}\label{secDiagrams}
%%%%%%%%%%%%%%%%%%%%%%%%%%%%%%%%%%%%%%%%%%%%%%%%%%
% FIGURE 6
\begin{figure*}[htb]
\begin{center}
\includegraphics[width=1\textwidth]{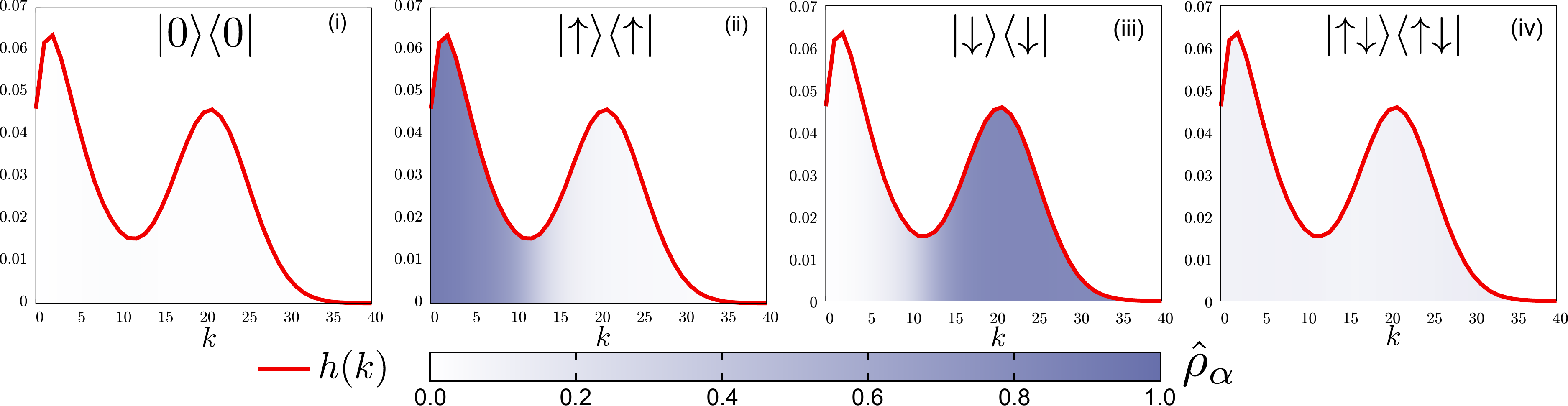}
\caption{(Color online) State-resolved density-matrix contribution
  to the expansion order histogram (see text). 
  Data are for the \tddpam with $\tpd\!=\!0.75$,  $\delta=0.10$ and
  bath spin-state $\sigma=\,\uparrow$. 
  The states of $\hat{\rho}_\alpha$ in different panels 
  are $|0\ket$ (i),  $|{\uparrow}\ket$ (ii), $|{\downarrow}\ket$ (iii),
  $|{\uparrow\downarrow}\ket$ (iv). 
  The figure shows the different contribution to the trace coming from the possible
  states configuration. 
  The empty and doubly occupied states  do not contribute much to the
  trace. The only significant contribution comes from the singly occupied
  states and interestingly the two peaks have complementary character.
}
\label{fig6}
\end{center}
\end{figure*}
%%%%%%%%%%%%%%%%%%%%%%%%%%%%%%%%%%%%%%%%%%%%%%%%%%
The previous analysis revealed the existence of a number of new
features in the spin susceptibility which are an inevitable
consequence of the inclusion of $p$-electrons. Yet,
$\chi_\mathrm{spin}$ is not the ideal physical quantity to understand
whether or not the new hybridization processes come entirely from $p$
degrees of freedom. In this section we introduce a quantity which
turns out to be able to discriminate between $d$ and $p$ character of
the hybridization processes

To begin with, we consider the hybridization function 
$\Delta_{dd}(\omega) = \sum_{l=1}^{N_b} V^2_l / (\omega^+ - \epsilon_l)$ on the
real-frequency axis. This contains essential information about the
formation of electronic excitations involved in the
screening channel. Then, using this discretized hybridization
function, we perform strong-coupling CTQMC calculations in order to
``visualize'' distinct classes of diagrams responsible for the
different  screening effects.

In Fig.\ref{fig4} we show the spectral density of the hybridization
function $\rho_0(\omega)=-\Im{\Delta_{dd}(\omega)}/\pi$ for several
values of $\tpd$ and finite doping. 
For the Hubbard model (see inset) this quantity has a finite
weight at the Fermi level, separated by higher energy feature
describing hybridization events with doubly occupied states (Hubbard band). 
For our model at tiny values of $\tpd$ $\rho_0(\omega)$ shows a
dramatic reduction of the weight at Fermi level, in agreement with the
loss of coherence of this metallic state.  It is very clear how
increasing $\tpd$ drives the formation of substantial
spectral weight below and at the Fermi level. Therefore the system gains a
lot of kinetic energy by introducing hybridization events in that frequency region. 

We now turn our attention to the effects introduced by  these ``new''
hybridization events, from a diagrammatic point of view. 
In Fig.\ref{fig5}  we show the fraction of diagrams
contributing to the fermionic trace in the CTQMC calculation as a
function of the expansion order, \ie the histogram $h(k)$ of the order of the
diagrams involved in the calculation.
The histogram of the ``$d$-only'' Hubbard model corresponds to a
single contribution near zero-order.   
For the \tddpam the low-expansion order feature 
gets instead less pronounced and, interestingly, a second structure
develops  at larger expansion orders. 
How can we understand this new higher-order peak and can we assign a ``label'' to it?

A first hint that the second peak at higher expansion orders reflects
the presence of  the $p$-orbital 
comes from the $\tpd$ dependence of its position.  
The expansion order histograms are very well 
fitted by a double gaussian function. 
The mean value of the high-order feature scales quadratically with
$\tpd$  (see inset of \figu{fig5}).

A more quantitative label is however needed. 
This is obtained by looking at the orbital, spin and expansion order
resolved site-reduced density matrix which can be directly
measured within the CTQMC calculation. 
The density matrix itself, whose diagonals are the 
state weights\cite{Werner2006PRB,Werner2007PRL}, is defined as 
$\hat{\rho}_\alpha=|{\alpha}\ket\bra{\alpha}|$
where $|{\alpha}\ket$ is an atomic many-body state of the local part of
the impurity Hamiltonian. In the present, simple, case of
density-density interaction we have
$|{\alpha}\ket=|{0}\ket$, $|{\uparrow}\ket$, $|{\downarrow}\ket$,
$|{\uparrow\downarrow}\ket$ and only the state weights are non-zero.
The density-matrix $\hat{\rho}_\alpha$ was previously
used in similar contexts, \eg Ref.~\onlinecite{Kunes2012PRL}, to obtain information about how much
time the system spends in a given local state. 
By resolving its measurement also in the expansion order, this quantity
tells us the probability to find the system in a certain atomic state
when there are a specific number of hybridization events with a given
spin and  orbital state.

With this piece of information we can assign an expansion-order
dependent  intensity to each histogram of Fig.\ref{fig5}. In
other words we look at the expansion order for a certain spin and orbital
and show as color intensity the value of the state weight for a fixed atomic
state at each expansion order. In Fig.\ref{fig6} we show this quantity
for the case  of $\tpd\!=\!0.75$. 
This unveils a very interesting property of the
hybridization-expansion CTQMC histogram for the \tddpam with 
finite hybridization.  

In the Hubbard model case the
order-resolved density matrix does not display a particularly strong
expansion-order dependence.
Therefore plotting the histogram with the colors from the density
matrix  would not be particularly revealing.  
Instead, in the case of the \tddpam the two peaks in
the histogram are characterized by almost completely
separated  classes of diagrams, as indicated by the complementary
color intensities in panels ($ii$) and ($iii$) of Fig.\ref{fig6}.
The density matrix used for the colors of Fig.\ref{fig6} is
calculated for a  fixed number $k$ of pairs of operators with spin
$\sigma=\uparrow$  in the local trace. 
Panel ($iii$) showing $\hat{\rho}_{\downarrow}$ therefore tells us that a
large number of diagrams with many (\ie high expansion-order)
spin-$\uparrow$  electrons hopping from and to the impurity contribute
to a measure of the local state  $|{\downarrow}\rangle$. 
This means that the impurity often visits the  state
$|{\uparrow\downarrow}\rangle$,  \ie the hybridization with the bath
makes  it often doubly occupied. Since $\tpd$ is large and the
$p$-band is  almost filled ($\bra n_p\ket\simeq$1.8 electrons), the $p$-orbital
acts as a very efficient particle-donor with respect to the impurity,
indicating that the peak at large $k$ describes hybridization
processes of mostly $p$-character. 
We have checked that in a specular situation, with the $p$-orbital
almost empty, the peak at large $k$ has intense color for
$\rho_{\uparrow}$, \ie  it corresponds to diagrams ``emptying'' the
impurity.  Since in that case the $p$-orbital ``accepts'' electrons
the same  conclusion of the large-$k$ peak being mostly of $p$-character holds.

%Kinetic energy:
The interpretation of the large-$k$ peak as stemming mainly from the
hybridization with the $p$-orbital suggests the following consideration.
Since in the hybridization-expansion CTQMC the mean value of the expansion
order histogram is proportional to the kinetic energy
\cite{Haule2007PRB},  for $\tpd\! \neq \! 0$ we can identify
the  presence of two distinct components in the system, one with
smaller kinetic energy predominantly of $d$-character and a more
mobile one of $p$-character (see Figs.\ref{fig5} and \ref{fig6}). 
The latter component is characterized by a large expansion order
therefore it corresponds to large hybridization strength. 

%Relation between chi and CTMQC
Let us note that we cannot directly relate peaks in the
expansion-order histogram to specific frequency structures of 
$\chi_\mathrm{spin}(\omega)$. 
Nevertheless, the previous analysis of the expansion-order resolved
density-matrix allowed us to indirectly relate the presence of the
feature at large $k$, containing  contributions to the
impurity screening coming from mainly $p$-electrons, to the 
$d$-$p$ character of the spin susceptibility. 
A more formal connection between the expansion-order histogram and response
functions of the impurity model can be established by evaluating higher
moments of the distribution. For example the width of the second
peak in $h(k)$ can give information about ``$p$-only'' contributions
to impurity susceptibilities. The present study provides a basis for
such an analysis, which we leave for a future investigation.

\section{Conclusions}\label{secConclusions}
In this paper, using DMFT we have investigated a simplified, yet generic,
model for $d$- (or $f$-) orbital materials, explicitly including
hybridization with more itinerant, \eg ligand $p$-, orbitals. We
focused on the paradigmatic example of doped charge-transfer
insulators.
In particular, we studied the evolution of the dynamical spin response
 $\chi_\mathrm{spin}(\omega)$ as a function of the hybridization.  
We pointed out the existence of different exchange mechanisms involved
in the  local moment screening. 
We showed that the direct exchange between $d$-orbitals, which
characterize  the screening physics of Hubbard-like models,  competes
with  indirect exchange mechanisms (Kondo singlet formation) involving 
hybridization  with conduction band electrons. 
We show how the presence of such different exchange mechanisms is reflected in the
structure of the spin susceptibility. The low-frequency feature  
associated to the metallic screening of local moments of the Hubbard 
model, acquires a multi-peaked structures containing contributions
from  both direct and indirect processes in the hybridized system. 
Moreover, the presence of additional screening channels is mirrored in the 
dynamical spin response by the formation of spectral weight at 
intermediate energies, extending over an energy range of the order
of the conduction electrons bandwidth. 
Using CTQMC, we characterized the
different  processes involved in the dynamical screening of
instantaneous local moments in terms of diagrams in the
hybridization-expansion around the atomic limit. 
We show that in the presence of finite hybridization, the
expansion-order  histogram acquires a characteristic double-peak
structure  revealing the concomitant presence 
of a more localized and a more mobile electronic component. 
A special analysis of the expansion-order state-resolved density
matrix allows us to assign a meaning to the peaks in the CTQMC
histogram, associating them to $d$- or $p$-hybridization events separately.
Our approach can be very useful in realistic calculations for quantifying the 
relative importance and the degree of intertwinement of the different
screening  channels of local moments in materials with 
orbitals of different degree of localization such, \eg metallic
cobaltates \cite{Kunes2012PRL}.

\paragraph*{Acknowledgments.}
We acknowledge useful discussions with S.~Ciuchi, L.~de' Medici,
V.~Hinkov,  J.~Kune\v{s} and A.~Toschi. 
G.S. acknowledges support by the Deutsche Forschungsgemeinschaft (FOR 1162).
A.A. and M.C. acknowledge financial support from the European Research Council under
FP7 Starting Independent Research Grant n.240524 ``SUPERBAD".

\bibliography{fullbib,localbib}

%merlin.mbs apsrev4-1.bst 2010-07-25 4.21a (PWD, AO, DPC) hacked
%Control: key (0)
%Control: author (0) dotless jnrlst
%Control: editor formatted (1) identically to author
%Control: production of article title (0) allowed
%Control: page (1) range
%Control: year (0) verbatim
%Control: production of eprint (0) enabled
\begin{thebibliography}{38}%
\makeatletter
\providecommand \@ifxundefined [1]{%
 \@ifx{#1\undefined}
}%
\providecommand \@ifnum [1]{%
 \ifnum #1\expandafter \@firstoftwo
 \else \expandafter \@secondoftwo
 \fi
}%
\providecommand \@ifx [1]{%
 \ifx #1\expandafter \@firstoftwo
 \else \expandafter \@secondoftwo
 \fi
}%
\providecommand \natexlab [1]{#1}%
\providecommand \enquote  [1]{``#1''}%
\providecommand \bibnamefont  [1]{#1}%
\providecommand \bibfnamefont [1]{#1}%
\providecommand \citenamefont [1]{#1}%
\providecommand \href@noop [0]{\@secondoftwo}%
\providecommand \href [0]{\begingroup \@sanitize@url \@href}%
\providecommand \@href[1]{\@@startlink{#1}\@@href}%
\providecommand \@@href[1]{\endgroup#1\@@endlink}%
\providecommand \@sanitize@url [0]{\catcode `\\12\catcode `\$12\catcode
  `\&12\catcode `\#12\catcode `\^12\catcode `\_12\catcode `\%12\relax}%
\providecommand \@@startlink[1]{}%
\providecommand \@@endlink[0]{}%
\providecommand \url  [0]{\begingroup\@sanitize@url \@url }%
\providecommand \@url [1]{\endgroup\@href {#1}{\urlprefix }}%
\providecommand \urlprefix  [0]{URL }%
\providecommand \Eprint [0]{\href }%
\providecommand \doibase [0]{http://dx.doi.org/}%
\providecommand \selectlanguage [0]{\@gobble}%
\providecommand \bibinfo  [0]{\@secondoftwo}%
\providecommand \bibfield  [0]{\@secondoftwo}%
\providecommand \translation [1]{[#1]}%
\providecommand \BibitemOpen [0]{}%
\providecommand \bibitemStop [0]{}%
\providecommand \bibitemNoStop [0]{.\EOS\space}%
\providecommand \EOS [0]{\spacefactor3000\relax}%
\providecommand \BibitemShut  [1]{\csname bibitem#1\endcsname}%
\let\auto@bib@innerbib\@empty
%</preamble>
\bibitem [{\citenamefont {Ament}\ \emph {et~al.}(2011)\citenamefont {Ament},
  \citenamefont {van Veenendaal}, \citenamefont {Devereaux}, \citenamefont
  {Hill},\ and\ \citenamefont {van~den Brink}}]{Ament2011}%
  \BibitemOpen
  \bibfield  {author} {\bibinfo {author} {\bibfnamefont {Luuk~J.P.}\
  \bibnamefont {Ament}}, \bibinfo {author} {\bibfnamefont {M.}~\bibnamefont
  {van Veenendaal}}, \bibinfo {author} {\bibfnamefont {T.P.}\ \bibnamefont
  {Devereaux}}, \bibinfo {author} {\bibfnamefont {J.P.}\ \bibnamefont {Hill}},
  \ and\ \bibinfo {author} {\bibfnamefont {J.}~\bibnamefont {van~den Brink}},\
  }\bibfield  {title} {\enquote {\bibinfo {title} {{Resonant inelastic x-ray
  scattering studies of elementary excitations}},}\ }\href {\doibase
  10.1103/RevModPhys.83.705} {\bibfield  {journal} {\bibinfo  {journal} {Rev.
  Mod. Phys.}\ }\textbf {\bibinfo {volume} {83}},\ \bibinfo {pages} {705--767}
  (\bibinfo {year} {2011})}\BibitemShut {NoStop}%
\bibitem [{\citenamefont {Braicovich}\ \emph {et~al.}(2010)\citenamefont
  {Braicovich}, \citenamefont {van~den Brink}, \citenamefont {Bisogni},
  \citenamefont {Sala}, \citenamefont {Ament}, \citenamefont {Brookes},
  \citenamefont {{De Luca}}, \citenamefont {Salluzzo}, \citenamefont {Schmitt},
  \citenamefont {Strocov},\ and\ \citenamefont
  {Ghiringhelli}}]{Braicovich2010}%
  \BibitemOpen
  \bibfield  {author} {\bibinfo {author} {\bibfnamefont {L.}~\bibnamefont
  {Braicovich}}, \bibinfo {author} {\bibfnamefont {J.}~\bibnamefont {van~den
  Brink}}, \bibinfo {author} {\bibfnamefont {V.}~\bibnamefont {Bisogni}},
  \bibinfo {author} {\bibfnamefont {M.~Moretti}\ \bibnamefont {Sala}}, \bibinfo
  {author} {\bibfnamefont {L.~J.~P.}\ \bibnamefont {Ament}}, \bibinfo {author}
  {\bibfnamefont {N.~B.}\ \bibnamefont {Brookes}}, \bibinfo {author}
  {\bibfnamefont {G.~M.}\ \bibnamefont {{De Luca}}}, \bibinfo {author}
  {\bibfnamefont {M.}~\bibnamefont {Salluzzo}}, \bibinfo {author}
  {\bibfnamefont {T.}~\bibnamefont {Schmitt}}, \bibinfo {author} {\bibfnamefont
  {V.~N.}\ \bibnamefont {Strocov}}, \ and\ \bibinfo {author} {\bibfnamefont
  {G.}~\bibnamefont {Ghiringhelli}},\ }\bibfield  {title} {\enquote {\bibinfo
  {title} {{Magnetic Excitations and Phase Separation in the Underdoped
  La\_\{2-x\}Sr\_\{x\}CuO\_\{4\} Superconductor Measured by Resonant Inelastic
  X-Ray Scattering}},}\ }\href {\doibase 10.1103/PhysRevLett.104.077002}
  {\bibfield  {journal} {\bibinfo  {journal} {Phys. Rev. Lett.}\ }\textbf
  {\bibinfo {volume} {104}},\ \bibinfo {pages} {077002} (\bibinfo {year}
  {2010})}\BibitemShut {NoStop}%
\bibitem [{\citenamefont {{Le Tacon}}\ \emph {et~al.}(2011)\citenamefont {{Le
  Tacon}}, \citenamefont {Ghiringhelli}, \citenamefont {Chaloupka},
  \citenamefont {Sala}, \citenamefont {Hinkov}, \citenamefont {Haverkort},
  \citenamefont {Minola}, \citenamefont {Bakr}, \citenamefont {Zhou},
  \citenamefont {Blanco-Canosa}, \citenamefont {Monney}, \citenamefont {Song},
  \citenamefont {Sun}, \citenamefont {Lin}, \citenamefont {{De Luca}},
  \citenamefont {Salluzzo}, \citenamefont {Khaliullin}, \citenamefont
  {Schmitt}, \citenamefont {Braicovich},\ and\ \citenamefont
  {Keimer}}]{Tacon2011}%
  \BibitemOpen
  \bibfield  {author} {\bibinfo {author} {\bibfnamefont {M.}~\bibnamefont {{Le
  Tacon}}}, \bibinfo {author} {\bibfnamefont {G.}~\bibnamefont {Ghiringhelli}},
  \bibinfo {author} {\bibfnamefont {J.}~\bibnamefont {Chaloupka}}, \bibinfo
  {author} {\bibfnamefont {M.~Moretti}\ \bibnamefont {Sala}}, \bibinfo {author}
  {\bibfnamefont {V.}~\bibnamefont {Hinkov}}, \bibinfo {author} {\bibfnamefont
  {M.~W.}\ \bibnamefont {Haverkort}}, \bibinfo {author} {\bibfnamefont
  {M.}~\bibnamefont {Minola}}, \bibinfo {author} {\bibfnamefont
  {M.}~\bibnamefont {Bakr}}, \bibinfo {author} {\bibfnamefont {K.~J.}\
  \bibnamefont {Zhou}}, \bibinfo {author} {\bibfnamefont {S.}~\bibnamefont
  {Blanco-Canosa}}, \bibinfo {author} {\bibfnamefont {C.}~\bibnamefont
  {Monney}}, \bibinfo {author} {\bibfnamefont {Y.~T.}\ \bibnamefont {Song}},
  \bibinfo {author} {\bibfnamefont {G.~L.}\ \bibnamefont {Sun}}, \bibinfo
  {author} {\bibfnamefont {C.~T.}\ \bibnamefont {Lin}}, \bibinfo {author}
  {\bibfnamefont {G.~M.}\ \bibnamefont {{De Luca}}}, \bibinfo {author}
  {\bibfnamefont {M.}~\bibnamefont {Salluzzo}}, \bibinfo {author}
  {\bibfnamefont {G.}~\bibnamefont {Khaliullin}}, \bibinfo {author}
  {\bibfnamefont {T.}~\bibnamefont {Schmitt}}, \bibinfo {author} {\bibfnamefont
  {L.}~\bibnamefont {Braicovich}}, \ and\ \bibinfo {author} {\bibfnamefont
  {B.}~\bibnamefont {Keimer}},\ }\bibfield  {title} {\enquote {\bibinfo {title}
  {{Intense paramagnon excitations in a large family of high-temperature
  superconductors}},}\ }\href {\doibase 10.1038/nphys2041} {\bibfield
  {journal} {\bibinfo  {journal} {Nat. Phys.}\ }\textbf {\bibinfo {volume}
  {7}},\ \bibinfo {pages} {725--730} (\bibinfo {year} {2011})}\BibitemShut
  {NoStop}%
\bibitem [{\citenamefont {Kroll}\ \emph {et~al.}(2006)\citenamefont {Kroll},
  \citenamefont {Knupfer}, \citenamefont {Geck}, \citenamefont {Hess},
  \citenamefont {Schwieger}, \citenamefont {Krabbes}, \citenamefont {Sekar},
  \citenamefont {Batchelor}, \citenamefont {Berger},\ and\ \citenamefont
  {B{\"u}chner}}]{Kroll2006PRB}%
  \BibitemOpen
  \bibfield  {author} {\bibinfo {author} {\bibfnamefont {T.}~\bibnamefont
  {Kroll}}, \bibinfo {author} {\bibfnamefont {M.}~\bibnamefont {Knupfer}},
  \bibinfo {author} {\bibfnamefont {J.}~\bibnamefont {Geck}}, \bibinfo {author}
  {\bibfnamefont {C.}~\bibnamefont {Hess}}, \bibinfo {author} {\bibfnamefont
  {T.}~\bibnamefont {Schwieger}}, \bibinfo {author} {\bibfnamefont
  {G.}~\bibnamefont {Krabbes}}, \bibinfo {author} {\bibfnamefont
  {C.}~\bibnamefont {Sekar}}, \bibinfo {author} {\bibfnamefont {D.~R.}\
  \bibnamefont {Batchelor}}, \bibinfo {author} {\bibfnamefont {H.}~\bibnamefont
  {Berger}}, \ and\ \bibinfo {author} {\bibfnamefont {B.}~\bibnamefont
  {B{\"u}chner}},\ }\bibfield  {title} {\enquote {\bibinfo {title} {{X-ray
  absorption spectroscopy of ${\mathrm{Na}}_{x}\mathrm{Co}{\mathrm{O}}_{2}$
  layered cobaltates}},}\ }\href {\doibase 10.1103/PhysRevB.74.115123}
  {\bibfield  {journal} {\bibinfo  {journal} {Phys. Rev. B}\ }\textbf {\bibinfo
  {volume} {74}},\ \bibinfo {pages} {115123} (\bibinfo {year}
  {2006})}\BibitemShut {NoStop}%
\bibitem [{\citenamefont {Lang}\ \emph {et~al.}(2008)\citenamefont {Lang},
  \citenamefont {Bobroff}, \citenamefont {Alloul}, \citenamefont {Collin},\
  and\ \citenamefont {Blanchard}}]{Lang2008PRB}%
  \BibitemOpen
  \bibfield  {author} {\bibinfo {author} {\bibfnamefont {G.}~\bibnamefont
  {Lang}}, \bibinfo {author} {\bibfnamefont {J.}~\bibnamefont {Bobroff}},
  \bibinfo {author} {\bibfnamefont {H.}~\bibnamefont {Alloul}}, \bibinfo
  {author} {\bibfnamefont {G.}~\bibnamefont {Collin}}, \ and\ \bibinfo {author}
  {\bibfnamefont {N.}~\bibnamefont {Blanchard}},\ }\bibfield  {title} {\enquote
  {\bibinfo {title} {{Spin correlations and cobalt charge states: Phase diagram
  of sodium cobaltates}},}\ }\href {\doibase 10.1103/PhysRevB.78.155116}
  {\bibfield  {journal} {\bibinfo  {journal} {Phys. Rev. B}\ }\textbf {\bibinfo
  {volume} {78}},\ \bibinfo {pages} {155116} (\bibinfo {year}
  {2008})}\BibitemShut {NoStop}%
\bibitem [{\citenamefont {Chi}\ \emph {et~al.}(2009)\citenamefont {Chi},
  \citenamefont {Schneidewind}, \citenamefont {Zhao}, \citenamefont {Harriger},
  \citenamefont {Li}, \citenamefont {Luo}, \citenamefont {Cao}, \citenamefont
  {Xu}, \citenamefont {Loewenhaupt}, \citenamefont {Hu},\ and\ \citenamefont
  {Dai}}]{Chi2009PRL}%
  \BibitemOpen
  \bibfield  {author} {\bibinfo {author} {\bibfnamefont {S.}~\bibnamefont
  {Chi}}, \bibinfo {author} {\bibfnamefont {A.}~\bibnamefont {Schneidewind}},
  \bibinfo {author} {\bibfnamefont {J.}~\bibnamefont {Zhao}}, \bibinfo {author}
  {\bibfnamefont {L.W.}\ \bibnamefont {Harriger}}, \bibinfo {author}
  {\bibfnamefont {L.}~\bibnamefont {Li}}, \bibinfo {author} {\bibfnamefont
  {Y.}~\bibnamefont {Luo}}, \bibinfo {author} {\bibfnamefont {G.}~\bibnamefont
  {Cao}}, \bibinfo {author} {\bibfnamefont {Z.}~\bibnamefont {Xu}}, \bibinfo
  {author} {\bibfnamefont {M.}~\bibnamefont {Loewenhaupt}}, \bibinfo {author}
  {\bibfnamefont {J.}~\bibnamefont {Hu}}, \ and\ \bibinfo {author}
  {\bibfnamefont {P.}~\bibnamefont {Dai}},\ }\bibfield  {title} {\enquote
  {\bibinfo {title} {{Inelastic Neutron-Scattering Measurements of a
  Three-Dimensional Spin Resonance in the FeAs-Based
  ${\mathrm{BaFe}}_{1.9}{\mathrm{Ni}}_{0.1}{\mathrm{As}}_{2}$
  Superconductor}},}\ }\href {\doibase 10.1103/PhysRevLett.102.107006}
  {\bibfield  {journal} {\bibinfo  {journal} {Phys. Rev. Lett.}\ }\textbf
  {\bibinfo {volume} {102}},\ \bibinfo {pages} {107006} (\bibinfo {year}
  {2009})}\BibitemShut {NoStop}%
\bibitem [{\citenamefont {Liu}\ \emph {et~al.}(2012)\citenamefont {Liu},
  \citenamefont {Harriger}, \citenamefont {Luo}, \citenamefont {Wang},
  \citenamefont {Ewings}, \citenamefont {Guidi}, \citenamefont {Park},
  \citenamefont {Haule}, \citenamefont {Kotliar}, \citenamefont {Hayden},\ and\
  \citenamefont {Dai}}]{Liu2012NP}%
  \BibitemOpen
  \bibfield  {author} {\bibinfo {author} {\bibfnamefont {M.}~\bibnamefont
  {Liu}}, \bibinfo {author} {\bibfnamefont {L.W.}\ \bibnamefont {Harriger}},
  \bibinfo {author} {\bibfnamefont {H.}~\bibnamefont {Luo}}, \bibinfo {author}
  {\bibfnamefont {M.}~\bibnamefont {Wang}}, \bibinfo {author} {\bibfnamefont
  {R.A.}\ \bibnamefont {Ewings}}, \bibinfo {author} {\bibfnamefont
  {T.}~\bibnamefont {Guidi}}, \bibinfo {author} {\bibfnamefont
  {H.}~\bibnamefont {Park}}, \bibinfo {author} {\bibfnamefont {K.}~\bibnamefont
  {Haule}}, \bibinfo {author} {\bibfnamefont {G.}~\bibnamefont {Kotliar}},
  \bibinfo {author} {\bibfnamefont {S~M.}\ \bibnamefont {Hayden}}, \ and\
  \bibinfo {author} {\bibfnamefont {P.}~\bibnamefont {Dai}},\ }\bibfield
  {title} {\enquote {\bibinfo {title} {{Nature of magnetic excitations in
  superconducting BaFe1.9Ni0.1As2}},}\ }\href
  {http://dx.doi.org/10.1038/nphys2268} {\bibfield  {journal} {\bibinfo
  {journal} {Nat. Phys.}\ }\textbf {\bibinfo {volume} {8}},\ \bibinfo {pages}
  {376--381} (\bibinfo {year} {2012})}\BibitemShut {NoStop}%
\bibitem [{\citenamefont {Zhou}\ \emph {et~al.}(2013)\citenamefont {Zhou},
  \citenamefont {Huang}, \citenamefont {Monney}, \citenamefont {Dai},
  \citenamefont {Strocov}, \citenamefont {Wang}, \citenamefont {Chen},
  \citenamefont {Zhang}, \citenamefont {Dai}, \citenamefont {Patthey},
  \citenamefont {van~den Brink}, \citenamefont {Ding},\ and\ \citenamefont
  {Schmitt}}]{Zhou2013NC}%
  \BibitemOpen
  \bibfield  {author} {\bibinfo {author} {\bibfnamefont {K.-J.}\ \bibnamefont
  {Zhou}}, \bibinfo {author} {\bibfnamefont {Y.-B.}\ \bibnamefont {Huang}},
  \bibinfo {author} {\bibfnamefont {C.}~\bibnamefont {Monney}}, \bibinfo
  {author} {\bibfnamefont {X.}~\bibnamefont {Dai}}, \bibinfo {author}
  {\bibfnamefont {V.~N.}\ \bibnamefont {Strocov}}, \bibinfo {author}
  {\bibfnamefont {N.-L.}\ \bibnamefont {Wang}}, \bibinfo {author}
  {\bibfnamefont {Z.-G.}\ \bibnamefont {Chen}}, \bibinfo {author}
  {\bibfnamefont {C.}~\bibnamefont {Zhang}}, \bibinfo {author} {\bibfnamefont
  {P.}~\bibnamefont {Dai}}, \bibinfo {author} {\bibfnamefont {L.}~\bibnamefont
  {Patthey}}, \bibinfo {author} {\bibfnamefont {J.}~\bibnamefont {van~den
  Brink}}, \bibinfo {author} {\bibfnamefont {H.}~\bibnamefont {Ding}}, \ and\
  \bibinfo {author} {\bibfnamefont {T.}~\bibnamefont {Schmitt}},\ }\bibfield
  {title} {\enquote {\bibinfo {title} {{Persistent high-energy spin excitations
  in iron-pnictide superconductors}},}\ }\href
  {http://dx.doi.org/10.1038/ncomms2428} {\bibfield  {journal} {\bibinfo
  {journal} {Nat. Commun.}\ }\textbf {\bibinfo {volume} {4}},\ \bibinfo {pages}
  {1470} (\bibinfo {year} {2013})}\BibitemShut {NoStop}%
\bibitem [{\citenamefont {Raas}\ \emph {et~al.}(2009)\citenamefont {Raas},
  \citenamefont {Grete},\ and\ \citenamefont {Uhrig}}]{Raas2009}%
  \BibitemOpen
  \bibfield  {author} {\bibinfo {author} {\bibfnamefont {C.}~\bibnamefont
  {Raas}}, \bibinfo {author} {\bibfnamefont {P.}~\bibnamefont {Grete}}, \ and\
  \bibinfo {author} {\bibfnamefont {G.}~\bibnamefont {Uhrig}},\ }\bibfield
  {title} {\enquote {\bibinfo {title} {{Emergent Collective Modes and Kinks in
  Electronic Dispersions}},}\ }\href {\doibase 10.1103/PhysRevLett.102.076406}
  {\bibfield  {journal} {\bibinfo  {journal} {Phys. Rev. Lett.}\ }\textbf
  {\bibinfo {volume} {102}},\ \bibinfo {pages} {076406} (\bibinfo {year}
  {2009})}\BibitemShut {NoStop}%
\bibitem [{\citenamefont {Grete}\ \emph {et~al.}(2011)\citenamefont {Grete},
  \citenamefont {Schmitt}, \citenamefont {Raas}, \citenamefont {Anders},\ and\
  \citenamefont {Uhrig}}]{Grete2011}%
  \BibitemOpen
  \bibfield  {author} {\bibinfo {author} {\bibfnamefont {P.}~\bibnamefont
  {Grete}}, \bibinfo {author} {\bibfnamefont {S.}~\bibnamefont {Schmitt}},
  \bibinfo {author} {\bibfnamefont {C.}~\bibnamefont {Raas}}, \bibinfo {author}
  {\bibfnamefont {F.B.}\ \bibnamefont {Anders}}, \ and\ \bibinfo {author}
  {\bibfnamefont {G.}~\bibnamefont {Uhrig}},\ }\bibfield  {title} {\enquote
  {\bibinfo {title} {{Kinks in the electronic dispersion of the Hubbard model
  away from half filling}},}\ }\href {\doibase 10.1103/PhysRevB.84.205104}
  {\bibfield  {journal} {\bibinfo  {journal} {Phys. Rev. B.}\ }\textbf
  {\bibinfo {volume} {84}},\ \bibinfo {pages} {205104} (\bibinfo {year}
  {2011})}\BibitemShut {NoStop}%
\bibitem [{\citenamefont {Hansmann}\ \emph {et~al.}(2010)\citenamefont
  {Hansmann}, \citenamefont {Arita}, \citenamefont {Toschi}, \citenamefont
  {Sakai}, \citenamefont {Sangiovanni},\ and\ \citenamefont
  {Held}}]{Hansmann2010PRL}%
  \BibitemOpen
  \bibfield  {author} {\bibinfo {author} {\bibfnamefont {P.}~\bibnamefont
  {Hansmann}}, \bibinfo {author} {\bibfnamefont {R.}~\bibnamefont {Arita}},
  \bibinfo {author} {\bibfnamefont {A.}~\bibnamefont {Toschi}}, \bibinfo
  {author} {\bibfnamefont {S.}~\bibnamefont {Sakai}}, \bibinfo {author}
  {\bibfnamefont {G.}~\bibnamefont {Sangiovanni}}, \ and\ \bibinfo {author}
  {\bibfnamefont {K.}~\bibnamefont {Held}},\ }\bibfield  {title} {\enquote
  {\bibinfo {title} {{Dichotomy between Large Local and Small Ordered Magnetic
  Moments in Iron-Based Superconductors}},}\ }\href {\doibase
  10.1103/PhysRevLett.104.197002} {\bibfield  {journal} {\bibinfo  {journal}
  {Phys. Rev. Lett.}\ }\textbf {\bibinfo {volume} {104}},\ \bibinfo {pages}
  {197002} (\bibinfo {year} {2010})}\BibitemShut {NoStop}%
\bibitem [{\citenamefont {Toschi}\ \emph {et~al.}(2012)\citenamefont {Toschi},
  \citenamefont {Arita}, \citenamefont {Hansmann}, \citenamefont {Sangiovanni},
  \citenamefont {Held},\ and\ \citenamefont {a.~Toschi}}]{Toschi2012PRB}%
  \BibitemOpen
  \bibfield  {author} {\bibinfo {author} {\bibfnamefont {A.}~\bibnamefont
  {Toschi}}, \bibinfo {author} {\bibfnamefont {R.}~\bibnamefont {Arita}},
  \bibinfo {author} {\bibfnamefont {P.}~\bibnamefont {Hansmann}}, \bibinfo
  {author} {\bibfnamefont {G.}~\bibnamefont {Sangiovanni}}, \bibinfo {author}
  {\bibfnamefont {K.}~\bibnamefont {Held}}, \ and\ \bibinfo {author}
  {\bibnamefont {a.~Toschi}},\ }\bibfield  {title} {\enquote {\bibinfo {title}
  {{Quantum dynamical screening of the local magnetic moment in Fe-based
  superconductors}},}\ }\href {\doibase 10.1103/PhysRevB.86.064411} {\bibfield
  {journal} {\bibinfo  {journal} {Physical Review B}\ }\textbf {\bibinfo
  {volume} {86}},\ \bibinfo {pages} {64411} (\bibinfo {year}
  {2012})}\BibitemShut {NoStop}%
\bibitem [{\citenamefont {Imada}\ \emph {et~al.}(1998)\citenamefont {Imada},
  \citenamefont {Fujimori},\ and\ \citenamefont {Tokura}}]{Imada1998RMP}%
  \BibitemOpen
  \bibfield  {author} {\bibinfo {author} {\bibfnamefont {Masatoshi}\
  \bibnamefont {Imada}}, \bibinfo {author} {\bibfnamefont {Atsushi}\
  \bibnamefont {Fujimori}}, \ and\ \bibinfo {author} {\bibfnamefont
  {Yoshinori}\ \bibnamefont {Tokura}},\ }\bibfield  {title} {\enquote {\bibinfo
  {title} {{Metal-insulator transitions}},}\ }\href {\doibase
  10.1103/RevModPhys.70.1039} {\bibfield  {journal} {\bibinfo  {journal} {Rev.
  Mod. Phys.}\ }\textbf {\bibinfo {volume} {70}},\ \bibinfo {pages}
  {1039--1263} (\bibinfo {year} {1998})}\BibitemShut {NoStop}%
\bibitem [{\citenamefont {Georges}\ \emph {et~al.}(1996)\citenamefont
  {Georges}, \citenamefont {Krauth}, \citenamefont {Kotliar},\ and\
  \citenamefont {Rozenberg}}]{Georges1996RMP}%
  \BibitemOpen
  \bibfield  {author} {\bibinfo {author} {\bibfnamefont {A.}~\bibnamefont
  {Georges}}, \bibinfo {author} {\bibfnamefont {W.}~\bibnamefont {Krauth}},
  \bibinfo {author} {\bibfnamefont {G.}~\bibnamefont {Kotliar}}, \ and\
  \bibinfo {author} {\bibfnamefont {M.J.~J}\ \bibnamefont {Rozenberg}},\
  }\bibfield  {title} {\enquote {\bibinfo {title} {{Dynamical mean-field theory
  of strongly correlated fermion systems and the limit of infinite
  dimensions}},}\ }\href {\doibase 10.1103/RevModPhys.68.13} {\bibfield
  {journal} {\bibinfo  {journal} {Rev. Mod. Phys.}\ }\textbf {\bibinfo {volume}
  {68}},\ \bibinfo {pages} {13} (\bibinfo {year} {1996})}\BibitemShut {NoStop}%
\bibitem [{\citenamefont {de' Medici}\ \emph {et~al.}(2005)\citenamefont {de'
  Medici}, \citenamefont {Georges}, \citenamefont {Kotliar},\ and\
  \citenamefont {Biermann}}]{Medici2005PRL}%
  \BibitemOpen
  \bibfield  {author} {\bibinfo {author} {\bibfnamefont {L.}~\bibnamefont {de'
  Medici}}, \bibinfo {author} {\bibfnamefont {A.}~\bibnamefont {Georges}},
  \bibinfo {author} {\bibfnamefont {G.}~\bibnamefont {Kotliar}}, \ and\
  \bibinfo {author} {\bibfnamefont {S.}~\bibnamefont {Biermann}},\ }\bibfield
  {title} {\enquote {\bibinfo {title} {{Mott Transition and Kondo Screening in
  f-Electron Metals}},}\ }\href {\doibase 10.1103/PhysRevLett.95.066402}
  {\bibfield  {journal} {\bibinfo  {journal} {Phys. Rev. Lett.}\ }\textbf
  {\bibinfo {volume} {95}},\ \bibinfo {pages} {066402} (\bibinfo {year}
  {2005})}\BibitemShut {NoStop}%
\bibitem [{\citenamefont {Jarrell}\ \emph {et~al.}(1993)\citenamefont
  {Jarrell}, \citenamefont {Akhlaghpour},\ and\ \citenamefont
  {Pruschke}}]{Jarrell1993PRL}%
  \BibitemOpen
  \bibfield  {author} {\bibinfo {author} {\bibfnamefont {M.}~\bibnamefont
  {Jarrell}}, \bibinfo {author} {\bibfnamefont {H.}~\bibnamefont
  {Akhlaghpour}}, \ and\ \bibinfo {author} {\bibfnamefont {T.}~\bibnamefont
  {Pruschke}},\ }\bibfield  {title} {\enquote {\bibinfo {title} {{Periodic
  Anderson model in infinite dimensions}},}\ }\href {\doibase
  10.1103/PhysRevLett.70.1670} {\bibfield  {journal} {\bibinfo  {journal}
  {Phys. Rev. Lett.}\ }\textbf {\bibinfo {volume} {70}},\ \bibinfo {pages}
  {1670--1673} (\bibinfo {year} {1993})}\BibitemShut {NoStop}%
\bibitem [{\citenamefont {Jarrell}(1995)}]{Jarrell1995PRB}%
  \BibitemOpen
  \bibfield  {author} {\bibinfo {author} {\bibfnamefont {M.}~\bibnamefont
  {Jarrell}},\ }\bibfield  {title} {\enquote {\bibinfo {title} {{Symmetric
  periodic Anderson model in infinite dimensions}},}\ }\href {\doibase
  10.1103/PhysRevB.51.7429} {\bibfield  {journal} {\bibinfo  {journal} {Phys.
  Rev. B}\ }\textbf {\bibinfo {volume} {51}},\ \bibinfo {pages} {7429--7440}
  (\bibinfo {year} {1995})}\BibitemShut {NoStop}%
\bibitem [{\citenamefont {Pruschke}\ \emph {et~al.}(2000)\citenamefont
  {Pruschke}, \citenamefont {Bulla},\ and\ \citenamefont
  {Jarrell}}]{Pruschke2000}%
  \BibitemOpen
  \bibfield  {author} {\bibinfo {author} {\bibfnamefont {T.}~\bibnamefont
  {Pruschke}}, \bibinfo {author} {\bibfnamefont {R.}~\bibnamefont {Bulla}}, \
  and\ \bibinfo {author} {\bibfnamefont {M.}~\bibnamefont {Jarrell}},\
  }\bibfield  {title} {\enquote {\bibinfo {title} {{Low-energy scale of the
  periodic Anderson model}},}\ }\href
  {http://prb.aps.org/abstract/PRB/v61/i19/p12799\_1} {\bibfield  {journal}
  {\bibinfo  {journal} {Phys. Rev. B}\ }\textbf {\bibinfo {volume} {61}},\
  \bibinfo {pages} {799--809} (\bibinfo {year} {2000})}\BibitemShut {NoStop}%
\bibitem [{\citenamefont {Sordi}\ \emph {et~al.}(2007)\citenamefont {Sordi},
  \citenamefont {Amaricci},\ and\ \citenamefont {Rozenberg}}]{Sordi2007PRL}%
  \BibitemOpen
  \bibfield  {author} {\bibinfo {author} {\bibfnamefont {G.}~\bibnamefont
  {Sordi}}, \bibinfo {author} {\bibfnamefont {A.}~\bibnamefont {Amaricci}}, \
  and\ \bibinfo {author} {\bibfnamefont {M.J.}\ \bibnamefont {Rozenberg}},\
  }\bibfield  {title} {\enquote {\bibinfo {title} {{Metal-Insulator Transitions
  in the Periodic Anderson Model}},}\ }\href {\doibase
  10.1103/PhysRevLett.99.196403} {\bibfield  {journal} {\bibinfo  {journal}
  {Phys. Rev. Lett.}\ }\textbf {\bibinfo {volume} {99}},\ \bibinfo {pages}
  {196403} (\bibinfo {year} {2007})}\BibitemShut {NoStop}%
\bibitem [{\citenamefont {Lechermann}\ \emph {et~al.}(2006)\citenamefont
  {Lechermann}, \citenamefont {Georges}, \citenamefont {Poteryaev},
  \citenamefont {Biermann}, \citenamefont {Posternak}, \citenamefont
  {Yamasaki},\ and\ \citenamefont {Andersen}}]{Lechermann2006PRB}%
  \BibitemOpen
  \bibfield  {author} {\bibinfo {author} {\bibfnamefont {F.}~\bibnamefont
  {Lechermann}}, \bibinfo {author} {\bibfnamefont {A.}~\bibnamefont {Georges}},
  \bibinfo {author} {\bibfnamefont {A.}~\bibnamefont {Poteryaev}}, \bibinfo
  {author} {\bibfnamefont {S.}~\bibnamefont {Biermann}}, \bibinfo {author}
  {\bibfnamefont {M.}~\bibnamefont {Posternak}}, \bibinfo {author}
  {\bibfnamefont {A.}~\bibnamefont {Yamasaki}}, \ and\ \bibinfo {author}
  {\bibfnamefont {O.~K.}\ \bibnamefont {Andersen}},\ }\bibfield  {title}
  {\enquote {\bibinfo {title} {{Dynamical mean-field theory using Wannier
  functions: A flexible route to electronic structure calculations of strongly
  correlated materials}},}\ }\href {\doibase 10.1103/PhysRevB.74.125120}
  {\bibfield  {journal} {\bibinfo  {journal} {Phys. Rev. B}\ }\textbf {\bibinfo
  {volume} {74}},\ \bibinfo {pages} {125120} (\bibinfo {year}
  {2006})}\BibitemShut {NoStop}%
\bibitem [{\citenamefont {Han}\ \emph {et~al.}(2011)\citenamefont {Han},
  \citenamefont {Wang}, \citenamefont {Marianetti},\ and\ \citenamefont
  {Millis}}]{Han2011PRL}%
  \BibitemOpen
  \bibfield  {author} {\bibinfo {author} {\bibfnamefont {M.~J.}\ \bibnamefont
  {Han}}, \bibinfo {author} {\bibfnamefont {Xin}\ \bibnamefont {Wang}},
  \bibinfo {author} {\bibfnamefont {C.~A.}\ \bibnamefont {Marianetti}}, \ and\
  \bibinfo {author} {\bibfnamefont {A.~J.}\ \bibnamefont {Millis}},\ }\bibfield
   {title} {\enquote {\bibinfo {title} {{Dynamical Mean-Field Theory of
  Nickelate Superlattices}},}\ }\href {\doibase 10.1103/PhysRevLett.107.206804}
  {\bibfield  {journal} {\bibinfo  {journal} {Phys. Rev. Lett.}\ }\textbf
  {\bibinfo {volume} {107}},\ \bibinfo {pages} {206804} (\bibinfo {year}
  {2011})}\BibitemShut {NoStop}%
\bibitem [{\citenamefont {Parragh}\ \emph {et~al.}(2013)\citenamefont
  {Parragh}, \citenamefont {Sangiovanni}, \citenamefont {Hansmann},
  \citenamefont {Hummel}, \citenamefont {Held},\ and\ \citenamefont
  {Toschi}}]{Parragh2013a}%
  \BibitemOpen
  \bibfield  {author} {\bibinfo {author} {\bibfnamefont {N.}~\bibnamefont
  {Parragh}}, \bibinfo {author} {\bibfnamefont {G.}~\bibnamefont
  {Sangiovanni}}, \bibinfo {author} {\bibfnamefont {P.}~\bibnamefont
  {Hansmann}}, \bibinfo {author} {\bibfnamefont {S.}~\bibnamefont {Hummel}},
  \bibinfo {author} {\bibfnamefont {K.}~\bibnamefont {Held}}, \ and\ \bibinfo
  {author} {\bibfnamefont {A.}~\bibnamefont {Toschi}},\ }\bibfield  {title}
  {\enquote {\bibinfo {title} {{Effective crystal field and Fermi surface
  topology: a comparison of d- and dp-orbital models}},}\ }\href
  {http://arxiv.org/abs/1303.2099} {\bibfield  {journal} {\bibinfo  {journal}
  {arXiv:1303.2099}\ } (\bibinfo {year} {2013})}\BibitemShut {NoStop}%
\bibitem [{\citenamefont {Haule}\ \emph {et~al.}(2013)\citenamefont {Haule},
  \citenamefont {Birol},\ and\ \citenamefont {Kotliar}}]{Haule2013a}%
  \BibitemOpen
  \bibfield  {author} {\bibinfo {author} {\bibfnamefont {K.}~\bibnamefont
  {Haule}}, \bibinfo {author} {\bibfnamefont {T.}~\bibnamefont {Birol}}, \ and\
  \bibinfo {author} {\bibfnamefont {G.}~\bibnamefont {Kotliar}},\ }\bibfield
  {title} {\enquote {\bibinfo {title} {{Covalency in transition metal oxides
  within all-electron Dynamical Mean Field Theory}},}\ }\href
  {http://arxiv.org/abs/1310.1158} {\bibfield  {journal} {\bibinfo  {journal}
  {arXiv:1310.1158}\ } (\bibinfo {year} {2013})}\BibitemShut {NoStop}%
\bibitem [{\citenamefont {Caffarel}\ and\ \citenamefont
  {Krauth}(1994)}]{Caffarel1994a}%
  \BibitemOpen
  \bibfield  {author} {\bibinfo {author} {\bibfnamefont {M.}~\bibnamefont
  {Caffarel}}\ and\ \bibinfo {author} {\bibfnamefont {W.}~\bibnamefont
  {Krauth}},\ }\bibfield  {title} {\enquote {\bibinfo {title} {{Exact
  diagonalization approach to correlated fermions in infinite dimensions: Mott
  transition and superconductivity}},}\ }\href {\doibase
  10.1103/PhysRevLett.72.1545} {\bibfield  {journal} {\bibinfo  {journal}
  {Phys. Rev. Lett.}\ }\textbf {\bibinfo {volume} {72}},\ \bibinfo {pages}
  {1545--1548} (\bibinfo {year} {1994})}\BibitemShut {NoStop}%
\bibitem [{\citenamefont {Capone}\ \emph {et~al.}(2007)\citenamefont {Capone},
  \citenamefont {de' Medici},\ and\ \citenamefont {Georges}}]{Capone2007PRB}%
  \BibitemOpen
  \bibfield  {author} {\bibinfo {author} {\bibfnamefont {M.}~\bibnamefont
  {Capone}}, \bibinfo {author} {\bibfnamefont {L.}~\bibnamefont {de' Medici}},
  \ and\ \bibinfo {author} {\bibfnamefont {A.}~\bibnamefont {Georges}},\
  }\bibfield  {title} {\enquote {\bibinfo {title} {{Solving the dynamical
  mean-field theory at very low temperatures using the Lanczos exact
  diagonalization}},}\ }\href {\doibase 10.1103/PhysRevB.76.245116} {\bibfield
  {journal} {\bibinfo  {journal} {Phys. Rev. B}\ }\textbf {\bibinfo {volume}
  {76}},\ \bibinfo {pages} {245116} (\bibinfo {year} {2007})}\BibitemShut
  {NoStop}%
\bibitem [{\citenamefont {Weber}\ \emph {et~al.}(2012)\citenamefont {Weber},
  \citenamefont {Amaricci}, \citenamefont {Capone},\ and\ \citenamefont
  {Littlewood}}]{Weber2012PRB}%
  \BibitemOpen
  \bibfield  {author} {\bibinfo {author} {\bibfnamefont {C.}~\bibnamefont
  {Weber}}, \bibinfo {author} {\bibfnamefont {A.}~\bibnamefont {Amaricci}},
  \bibinfo {author} {\bibfnamefont {M.}~\bibnamefont {Capone}}, \ and\ \bibinfo
  {author} {\bibfnamefont {P.B.}\ \bibnamefont {Littlewood}},\ }\bibfield
  {title} {\enquote {\bibinfo {title} {{Augmented hybrid exact-diagonalization
  solver for dynamical mean field theory}},}\ }\href {\doibase
  10.1103/PhysRevB.86.115136} {\bibfield  {journal} {\bibinfo  {journal} {Phys.
  Rev. B}\ }\textbf {\bibinfo {volume} {86}},\ \bibinfo {pages} {1--5}
  (\bibinfo {year} {2012})}\BibitemShut {NoStop}%
\bibitem [{\citenamefont {Werner}\ \emph {et~al.}(2006)\citenamefont {Werner},
  \citenamefont {Comanac}, \citenamefont {de' Medici}, \citenamefont {Troyer},\
  and\ \citenamefont {Millis}}]{Werner2006PRL}%
  \BibitemOpen
  \bibfield  {author} {\bibinfo {author} {\bibfnamefont {P.}~\bibnamefont
  {Werner}}, \bibinfo {author} {\bibfnamefont {A.}~\bibnamefont {Comanac}},
  \bibinfo {author} {\bibfnamefont {L.}~\bibnamefont {de' Medici}}, \bibinfo
  {author} {\bibfnamefont {M.}~\bibnamefont {Troyer}}, \ and\ \bibinfo {author}
  {\bibfnamefont {A.~J.}\ \bibnamefont {Millis}},\ }\bibfield  {title}
  {\enquote {\bibinfo {title} {{Continuous-Time Solver for Quantum Impurity
  Models}},}\ }\href {\doibase 10.1103/PhysRevLett.97.076405} {\bibfield
  {journal} {\bibinfo  {journal} {Phys. Rev. Lett.}\ }\textbf {\bibinfo
  {volume} {97}},\ \bibinfo {pages} {76405} (\bibinfo {year}
  {2006})}\BibitemShut {NoStop}%
\bibitem [{\citenamefont {Haule}(2007)}]{Haule2007PRB}%
  \BibitemOpen
  \bibfield  {author} {\bibinfo {author} {\bibfnamefont {K.}~\bibnamefont
  {Haule}},\ }\bibfield  {title} {\enquote {\bibinfo {title} {{Quantum Monte
  Carlo impurity solver for cluster dynamical mean-field theory and electronic
  structure calculations with adjustable cluster base}},}\ }\href {\doibase
  10.1103/PhysRevB.75.155113} {\bibfield  {journal} {\bibinfo  {journal} {Phys.
  Rev. B}\ }\textbf {\bibinfo {volume} {75}},\ \bibinfo {pages} {155113}
  (\bibinfo {year} {2007})}\BibitemShut {NoStop}%
\bibitem [{\citenamefont {L\"{a}uchli}\ and\ \citenamefont
  {Werner}(2009)}]{Lauchli2009}%
  \BibitemOpen
  \bibfield  {author} {\bibinfo {author} {\bibfnamefont {A.~M.}\ \bibnamefont
  {L\"{a}uchli}}\ and\ \bibinfo {author} {\bibfnamefont {P.}~\bibnamefont
  {Werner}},\ }\bibfield  {title} {\enquote {\bibinfo {title} {{Krylov
  implementation of the hybridization expansion impurity solver and application
  to 5-orbital models}},}\ }\href {\doibase 10.1103/PhysRevB.80.235117}
  {\bibfield  {journal} {\bibinfo  {journal} {Phys. Rev. B}\ }\textbf {\bibinfo
  {volume} {80}},\ \bibinfo {pages} {1--8} (\bibinfo {year}
  {2009})}\BibitemShut {NoStop}%
\bibitem [{\citenamefont {Parragh}\ \emph {et~al.}(2012)\citenamefont
  {Parragh}, \citenamefont {Toschi}, \citenamefont {Held},\ and\ \citenamefont
  {Sangiovanni}}]{Parragh2012PRB}%
  \BibitemOpen
  \bibfield  {author} {\bibinfo {author} {\bibfnamefont {N.}~\bibnamefont
  {Parragh}}, \bibinfo {author} {\bibfnamefont {A.}~\bibnamefont {Toschi}},
  \bibinfo {author} {\bibfnamefont {K.}~\bibnamefont {Held}}, \ and\ \bibinfo
  {author} {\bibfnamefont {G.}~\bibnamefont {Sangiovanni}},\ }\bibfield
  {title} {\enquote {\bibinfo {title} {{Conserved quantities of SU(2)-invariant
  interactions for correlated fermions and the advantages for quantum Monte
  Carlo simulations}},}\ }\href {\doibase 10.1103/PhysRevB.86.155158}
  {\bibfield  {journal} {\bibinfo  {journal} {Phys. Rev. B}\ }\textbf {\bibinfo
  {volume} {86}},\ \bibinfo {pages} {155158} (\bibinfo {year}
  {2012})}\BibitemShut {NoStop}%
\bibitem [{\citenamefont {Tahvildar-Zadeh}\ \emph {et~al.}(1997)\citenamefont
  {Tahvildar-Zadeh}, \citenamefont {Jarrell},\ and\ \citenamefont
  {Freericks}}]{Tahvildar-Zadeh1997PRB}%
  \BibitemOpen
  \bibfield  {author} {\bibinfo {author} {\bibfnamefont {A.N.}\ \bibnamefont
  {Tahvildar-Zadeh}}, \bibinfo {author} {\bibfnamefont {M.}~\bibnamefont
  {Jarrell}}, \ and\ \bibinfo {author} {\bibfnamefont {J.K.}\ \bibnamefont
  {Freericks}},\ }\bibfield  {title} {\enquote {\bibinfo {title} {{Protracted
  screening in the periodic Anderson model}},}\ }\href {\doibase
  10.1103/PhysRevB.55.R3332} {\bibfield  {journal} {\bibinfo  {journal} {Phys.
  Rev. B}\ }\textbf {\bibinfo {volume} {55}},\ \bibinfo {pages} {R3332--R3335}
  (\bibinfo {year} {1997})}\BibitemShut {NoStop}%
\bibitem [{\citenamefont {Burdin}\ \emph {et~al.}(2000)\citenamefont {Burdin},
  \citenamefont {Georges},\ and\ \citenamefont {Grempel}}]{Burdin2000PRL}%
  \BibitemOpen
  \bibfield  {author} {\bibinfo {author} {\bibfnamefont {S.}~\bibnamefont
  {Burdin}}, \bibinfo {author} {\bibfnamefont {A.}~\bibnamefont {Georges}}, \
  and\ \bibinfo {author} {\bibfnamefont {D.}~\bibnamefont {Grempel}},\
  }\bibfield  {title} {\enquote {\bibinfo {title} {{Coherence Scale of the
  Kondo Lattice}},}\ }\href {\doibase 10.1103/PhysRevLett.85.1048} {\bibfield
  {journal} {\bibinfo  {journal} {Phys. Rev. Lett.}\ }\textbf {\bibinfo
  {volume} {85}},\ \bibinfo {pages} {1048--1051} (\bibinfo {year}
  {2000})}\BibitemShut {NoStop}%
\bibitem [{\citenamefont {Burdin}\ and\ \citenamefont
  {Zlati\'{c}}(2009)}]{Burdin2009PRB}%
  \BibitemOpen
  \bibfield  {author} {\bibinfo {author} {\bibfnamefont {S.}~\bibnamefont
  {Burdin}}\ and\ \bibinfo {author} {\bibfnamefont {V.}~\bibnamefont
  {Zlati\'{c}}},\ }\bibfield  {title} {\enquote {\bibinfo {title} {{Multiple
  temperature scales of the periodic Anderson model: Slave boson approach}},}\
  }\href {\doibase 10.1103/PhysRevB.79.115139} {\bibfield  {journal} {\bibinfo
  {journal} {Phys. Rev. B}\ }\textbf {\bibinfo {volume} {79}},\ \bibinfo
  {pages} {115139} (\bibinfo {year} {2009})}\BibitemShut {NoStop}%
\bibitem [{\citenamefont {Amaricci}\ \emph {et~al.}(2008)\citenamefont
  {Amaricci}, \citenamefont {Sordi},\ and\ \citenamefont
  {Rozenberg}}]{Amaricci2008PRL}%
  \BibitemOpen
  \bibfield  {author} {\bibinfo {author} {\bibfnamefont {A.}~\bibnamefont
  {Amaricci}}, \bibinfo {author} {\bibfnamefont {G.}~\bibnamefont {Sordi}}, \
  and\ \bibinfo {author} {\bibfnamefont {M.J.}\ \bibnamefont {Rozenberg}},\
  }\bibfield  {title} {\enquote {\bibinfo {title} {{Non-Fermi-Liquid Behavior
  in the Periodic Anderson Model}},}\ }\href {\doibase
  10.1103/PhysRevLett.101.146403} {\bibfield  {journal} {\bibinfo  {journal}
  {Phys. Rev. Lett.}\ }\textbf {\bibinfo {volume} {101}},\ \bibinfo {pages}
  {1--4} (\bibinfo {year} {2008})}\BibitemShut {NoStop}%
\bibitem [{\citenamefont {Amaricci}\ \emph {et~al.}(2012)\citenamefont
  {Amaricci}, \citenamefont {de' Medici}, \citenamefont {Sordi}, \citenamefont
  {Rozenberg},\ and\ \citenamefont {Capone}}]{Amaricci2012PRB}%
  \BibitemOpen
  \bibfield  {author} {\bibinfo {author} {\bibfnamefont {A.}~\bibnamefont
  {Amaricci}}, \bibinfo {author} {\bibfnamefont {L.}~\bibnamefont {de'
  Medici}}, \bibinfo {author} {\bibfnamefont {G.}~\bibnamefont {Sordi}},
  \bibinfo {author} {\bibfnamefont {M.~J}\ \bibnamefont {Rozenberg}}, \ and\
  \bibinfo {author} {\bibfnamefont {M.}~\bibnamefont {Capone}},\ }\bibfield
  {title} {\enquote {\bibinfo {title} {{Path to poor coherence in the periodic
  Anderson model from Mott physics and hybridization}},}\ }\href {\doibase
  10.1103/PhysRevB.85.235110} {\bibfield  {journal} {\bibinfo  {journal} {Phys.
  Rev. B}\ }\textbf {\bibinfo {volume} {85}},\ \bibinfo {pages} {235110}
  (\bibinfo {year} {2012})}\BibitemShut {NoStop}%
\bibitem [{\citenamefont {Werner}\ and\ \citenamefont
  {Millis}(2006)}]{Werner2006PRB}%
  \BibitemOpen
  \bibfield  {author} {\bibinfo {author} {\bibfnamefont {Philipp}\ \bibnamefont
  {Werner}}\ and\ \bibinfo {author} {\bibfnamefont {Andrew~J.}\ \bibnamefont
  {Millis}},\ }\bibfield  {title} {\enquote {\bibinfo {title} {{Hybridization
  expansion impurity solver: General formulation and application to Kondo
  lattice and two-orbital models}},}\ }\href {\doibase
  10.1103/PhysRevB.74.155107} {\bibfield  {journal} {\bibinfo  {journal} {Phys.
  Rev. B}\ }\textbf {\bibinfo {volume} {74}},\ \bibinfo {pages} {155107}
  (\bibinfo {year} {2006})}\BibitemShut {NoStop}%
\bibitem [{\citenamefont {Werner}\ and\ \citenamefont
  {Millis}(2007)}]{Werner2007PRL}%
  \BibitemOpen
  \bibfield  {author} {\bibinfo {author} {\bibfnamefont {P.}~\bibnamefont
  {Werner}}\ and\ \bibinfo {author} {\bibfnamefont {A.~J.J.}\ \bibnamefont
  {Millis}},\ }\bibfield  {title} {\enquote {\bibinfo {title} {{High-Spin to
  Low-Spin and Orbital Polarization Transitions in Multiorbital Mott
  Systems}},}\ }\href {\doibase 10.1103/PhysRevLett.99.126405} {\bibfield
  {journal} {\bibinfo  {journal} {Phys. Rev. Lett.}\ }\textbf {\bibinfo
  {volume} {99}},\ \bibinfo {pages} {126405} (\bibinfo {year}
  {2007})}\BibitemShut {NoStop}%
\bibitem [{\citenamefont {{Kune\ifmmode \check{s}\else \v{s}\fi{}}}\ \emph
  {et~al.}(2012)\citenamefont {{Kune\ifmmode \check{s}\else \v{s}\fi{}}},
  \citenamefont {{K\ifmmode \check{r}\else \v{r}\fi{}{\'a}pek}}, \citenamefont
  {Parragh}, \citenamefont {Sangiovanni}, \citenamefont {Toschi},\ and\
  \citenamefont {Kozhevnikov}}]{Kunes2012PRL}%
  \BibitemOpen
  \bibfield  {author} {\bibinfo {author} {\bibfnamefont {J.}~\bibnamefont
  {{Kune\ifmmode \check{s}\else \v{s}\fi{}}}}, \bibinfo {author} {\bibfnamefont
  {V.}~\bibnamefont {{K\ifmmode \check{r}\else \v{r}\fi{}{\'a}pek}}}, \bibinfo
  {author} {\bibfnamefont {N.}~\bibnamefont {Parragh}}, \bibinfo {author}
  {\bibfnamefont {G.}~\bibnamefont {Sangiovanni}}, \bibinfo {author}
  {\bibfnamefont {A.}~\bibnamefont {Toschi}}, \ and\ \bibinfo {author}
  {\bibfnamefont {A.~V.}\ \bibnamefont {Kozhevnikov}},\ }\bibfield  {title}
  {\enquote {\bibinfo {title} {{Spin State of Negative Charge-Transfer Material
  ${\mathrm{SrCoO}}_{3}$}},}\ }\href {\doibase 10.1103/PhysRevLett.109.117206}
  {\bibfield  {journal} {\bibinfo  {journal} {Phys. Rev. Lett.}\ }\textbf
  {\bibinfo {volume} {109}},\ \bibinfo {pages} {117206} (\bibinfo {year}
  {2012})}\BibitemShut {NoStop}%
\end{thebibliography}%

\end{document}